\documentclass[aps,prl,twocolumn,superscriptaddress,showpacs]{revtex4}
\usepackage{amsmath,amssymb,graphicx}

\begin{document}

\title{
Photo-induced demagnetization and polaron binding energy increase\\
observed by mid-infrared pump-probe spectroscopy in ferromagnetic Ga$_{0.94}$Mn$_{0.06}$As
}

\author{E. Kojima}
\affiliation{%
Department of Applied Physics, the University of Tokyo, and 
Solution Oriented Research for Science and Technology (SORST), JST, 7-3-1 Hongo, Bunkyo-ku, Tokyo 113-8656, Japan
}

\author{J.B. Heroux}
\affiliation{%
Department of Applied Physics, the University of Tokyo, and 
Solution Oriented Research for Science and Technology (SORST), JST, 7-3-1 Hongo, Bunkyo-ku, Tokyo 113-8656, Japan
}

\author{R. Shimano}
\altaffiliation{present address: Department of Physics, The University of Tokyo}
\affiliation{%
Department of Applied Physics, the University of Tokyo, and 
Solution Oriented Research for Science and Technology (SORST), JST, 7-3-1 Hongo, Bunkyo-ku, Tokyo 113-8656, Japan
}

\author{Y. Hashimoto}
\affiliation{%
Institute for Solid State Physics, The University of Tokyo,
5-1-5 Kashiwanoha, Chiba 277-8581, Japan
}

\author{S. Katsumoto}
\affiliation{%
Institute for Solid State Physics, The University of Tokyo,
5-1-5 Kashiwanoha, Chiba 277-8581, Japan
}

\author{Y. Iye}
\affiliation{%
Institute for Solid State Physics, The University of Tokyo,
5-1-5 Kashiwanoha, Chiba 277-8581, Japan
}

\author{M. Kuwata-Gonokami}
\email{gonokami@ap.t.u-tokyo.ac.jp}
\affiliation{%
Department of Applied Physics, the University of Tokyo, and 
Solution Oriented Research for Science and Technology (SORST), JST, 7-3-1 Hongo, Bunkyo-ku, Tokyo 113-8656, Japan
}
\date{\today}

\begin{abstract}

Time-resolved transmittance measurements performed on Ga$_{0.94}$Mn$_{0.06}$As in the vicinity of the Mn-induced mid-infrared absorption band are presented. Upon photo-excitation, a slow increase (hundreds of ps timescale) of the 
differential transmittance is observed and found to be directly related to demagnetization. The temporal profiles of the transmittance and of the demagnetization measured by time-resolved magneto-optical Kerr spectroscopy are found to coincide. Well below the Curie temperature, the maximum amplitude of the slow component of the differential transmittance as a function of the probe energy is on the rising edge of the linear absorption peak, suggesting that ferromagnetic ordering can be explained by a coupling of the Mn local spins through bound magnetic polarons.
\end{abstract}
\pacs{}

\maketitle

	The In$_{1-x}$Mn$_x$As \cite{Munekata} and Ga$_{1-x}$Mn$_x$As \cite{Ohno} dilute magnetic semiconductors form a new class of ferromagnetic materials in which spin-related physical properties could potentially be used to add new functionality in conventional, III-V-based electronic and optoelectronic devices. Even though it has been established that ferromagnetic order is mediated by holes in these materials, the detailed mechanism of spin interaction is still unclear\cite{Matsukura,Akai}. The sensitivity of the magnetization to the free carrier concentration suggests that a good understanding of those mechanisms could potentially lead to higher Curie temperatures, a requirement for device applications.

    A key experimental feature which could allow to better understand ferromagnetism in these semiconductors is the presence of a manganese-induced broad mid-infrared (MIR) absorption band with a peak located around 200 meV for Ga$_{1-x}$Mn$_x$As\cite{Singley,Hirakawa}. The presence of an absorption peak, as opposed to a Drude-like shape, indicates that unlike the case of an ideal doped semiconductor in which the midinfrared response exhibits a metallic behavior, charge carriers are localized and the optical response is mediated by doped carriers. This implies that acceptor-bound magnetic polarons (ABMP) are formed in which Mn spins are aligned around holes, which are localized through a hybridization between the As 4p and Mn 3d states\cite{Kasuya,Bugajski}. Hence polaron effects should be taken into account in physical models\cite{Kaminski}.

\begin{figure}
\includegraphics[width=\linewidth]{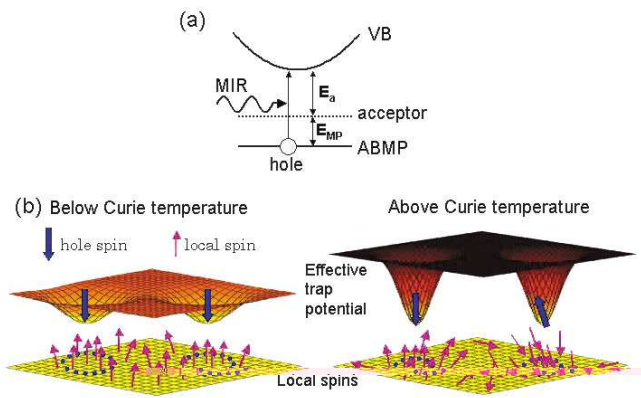}
\caption{
(a) Valence band configuration and midinfrared optical transition in Ga$_{1-x}$Mn$_x$As following the acceptor bound magnetic polaron (ABMP) model of carrier-induced ferromagnetism. (b) Schematic illustration of the spatial variation of hole energy and spin configurations below (left) and above (right) the Curie temperature according to the ABMP model. A deeper potential well (i.e., a larger binding energy) is obtained above the Curie temperature.
}
\label{fig3}
\end{figure}

    A Ga$_{1-x}$Mn$_x$As valence band diagram is schematically shown in Fig. \ref{fig3}(a). The Mn acceptor level is located at an energy $Ea\sim$110 meV away from the valence band\cite{Tarhan} and the position of this level is shifted by a value $E_{\rm MP}$ due to the $p$-$d$ exchange interaction. A spatial variation of hole energy and the hole and Mn spin orientations are shown in Fig. \ref{fig3}(b). Below the Curie temperature, the ABMPs are correlated and have the same orientation (left). As the temperature rises, the ABMPs progressively loose their alignment with each other. This fluctuation enhances the energy of itinerant carriers through the exchange interaction and the effective binding energy ($E_{\rm MP}$) increases.

    The correlation between midinfrared absorption properties and ferromagnetism has been previously examined experimentally. The absorption strength is strongly temperature-dependent below the Curie temperature, and the higher Curie temperature obtained due to sample annealing also leads to enhanced absorption. Hirakawa et al.\cite{Hirakawa} reported that the GaMnAs spectrum contained a temperature dependent component with Drude-like tailing. But the temperature dependence of the resistivity below the Curie temperature is very weak so that the variation in absorption strength cannot be simply attributed to a variation of the free carrier concentration. Mid-infrared absorption is related to ABMPs and other paramagnetic levels\cite{Ishiwata}, but the detailed mechanism remains unclear. To better understand the interaction between carriers and localized spins, extracting the component of the absorption band related to ferromagnetism is required.

    Ultrafast optical excitation is a powerful technique to investigate ferromagnetic materials because the spin, lattice and electron temperature rise-times caused by a light pulse are usually different\cite{Beaurepaire}. Previous time-resolved magneto-optical Kerr measurements (TR-MOKE) revealed a very long spin relaxation time in the hundreds of picoseconds in Ga$_{1-x}$Mn$_x$As due to the spin-dependent valence band structure of this material\cite{Kojima}. Because of this feature, a correlation exists between the optical response in the midinfrared and the magnetic behavior.

    In this work, we performed time-resolved mid-infrared transmittance measurements of a Ga$_{0.94}$Mn$_{0.06}$As epitaxial layer and were able to isolate and observe the ferromagnetism-related component of the mid-infrared absorption. In the hundreds of picoseconds timescale, the differential transmittance is directly related to magnetic properties and mid-infrared absorption involves an Mn-acceptor level. We also show that light-induced demagnetization causes a shift of the acceptor level (a blueshift of the linear mid-infrared absorption peak), strongly suggesting that ferromagnetic ordering is due to a coupling of the local spins of the Mn atoms with localized holes in agreement with the ABMP model.

    A 1.05 $\mu$m thick Ga$_{0.94}$Mn$_{0.06}$As layer was grown by molecular beam epitaxy on an undoped [001] GaAs substrate and annealed at 280C for 15 minutes. The Curie temperature of the sample was determined from static magnetization measured by a superconducting quantum interference device (SQUID) and found to be 110K. The pump-probe spectroscopy setup consisted of an optical parametric amplifier (OPA) and an amplified mode-locked Ti-sapphire laser emitting light pulses with a 1.55 eV photon energy and a 150 fs duration at a 1 kHz repetition rate. Frequency-doubled pulses with a 3.1 eV photon energy and a 660 ${\rm \mu}$J/cm$^2$ intensity were used as a pump. The probe pulses, having an energy ranging between 100 and 400 meV, were obtained by the difference frequency generation of an AgGaS$_2$ crystal using the signal and idler beams of the OPA. A liquid nitrogen cooled HgCdTe photodetector was used to detect the mid-infrared probe pulses transmitted through the sample, which was cooled in a closed-cycle cryostat.

\begin{figure}
\includegraphics[width=\linewidth]{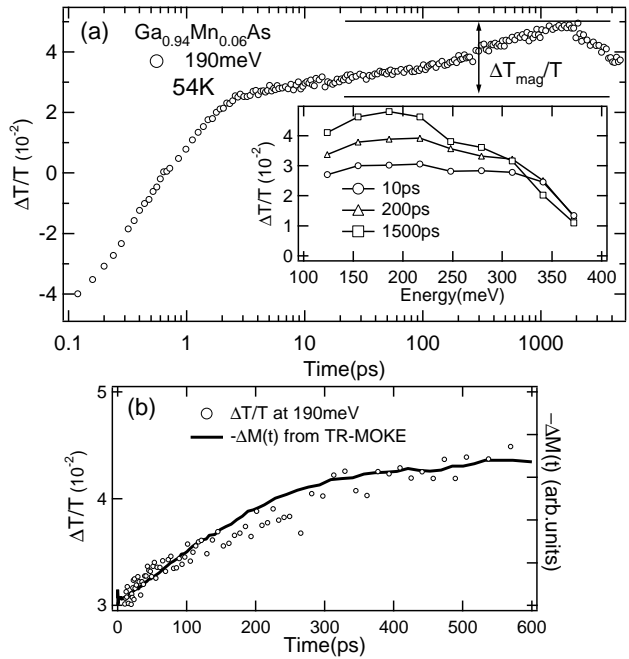}
\caption{
(a) Temporal evolution of the differential transmittance $\varDelta T/T$.
The magnitude of the spin-related slow component $\varDelta T/T_{\rm mag}$ is shown by horizontal lines. Inset shows the energy dependence of $\varDelta T/T$.
(b) Magnitude of the photo-induced demagnetization measured by time-resolved magneto-optical Kerr effect (right axis)
and of $\varDelta T/T$ (left axis).
}
\label{fig1}
\end{figure}

    A typical temporal evolution of the differential transmission at a 190 meV probe energy after photo-excitation is shown in Fig.\ref{fig1}(a) and can be divided into three distinct parts. Up to approximately 2 ps, the differential transmittance is governed by two competing processes: photo-induced increase of the free carrier population leading to a negative $\varDelta T$ through a change of the dielectric constant; and lattice and carrier heating leading to a positive $\varDelta T$. The observed shape of $\varDelta T/T$ is the sum of these two components. Above 2 ps approximately, non-radiative recombination of photo-generated carriers is completed, and there is a transition to a much slower rise of $\varDelta T$ reaching a maximum around 1500 ps. In this time interval, the transmittance is governed by a partial thermal isolation of carriers with different spins leading to a slow spin relaxation, which we previously described as a half-metallic electronic structure\cite{Kojima}. The amplitude of this slow increase, critical for our analysis, is denoted as $\varDelta T_{\rm mag}/T$ in Fig.\ref{fig1}. Above 1500 ps approximately, the
system begins to relax to its initial state. The inset shows the probe energy dependence of $\varDelta T/T$.

    In Fig.\ref{fig1}(b) $\varDelta T/T$ is plotted on a linear time scale from 10 to 600 ps along with the increase of demagnetization $-\varDelta M$ obtained by a two color TR-MOKE measurement\cite{Kojima} to show how the temporal response of the transmittance under pulsed photo-excitation is directly related to magnetization. Above 600 ps the demagnetization observed by the two techniques differs only due to the spatially inhomogeneous temperature rises within the material. This agreement between the two independent experimental techniques is not fortuitous since it is observed for a range of infrared probe energies around the midinfrared absorption peak. Even though $\varDelta T_{\rm mag}/T$ varies with the probe energy as can be inferred from the inset of Fig. 1 (a), the shape of the slow increase of $\varDelta T/T$ consistently matches the measured demagnetization.

\begin{figure}
\includegraphics[width=\linewidth]{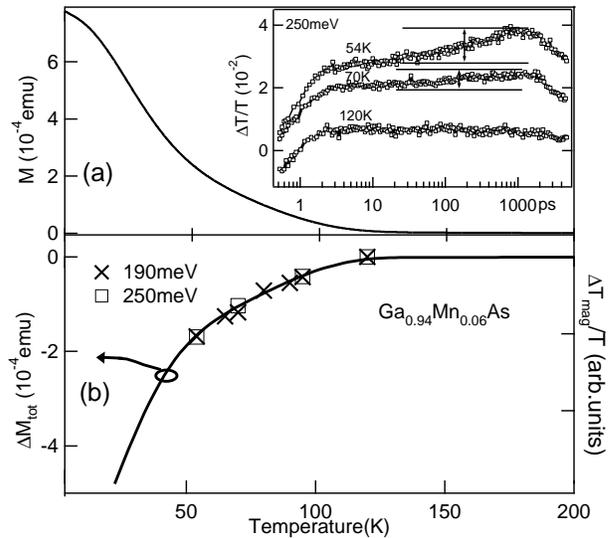}
\caption{
(a) Temperature dependence of the magnetization measured with a SQUID. Horizontal lines in the inset show the magnitude of the slow response of the differential transmittance $\varDelta T/T_{\rm mag}$. (b) The solid line shows the magnetization change due to a 660 ${\rm \mu J/cm^2}$ heat injection (see text). Circles and squares show $\varDelta T/T_{\rm mag}$ at 190 and 250 meV probe energies respectively (the right vertical axis). The scaling factors for the two sets of data are different.
}
\label{fig2}
\end{figure}

    We investigate next the variation of the transmittance as a function of the temperature ${\mathcal T}$. The magnetization $M$ of the sample measured by SQUID with a 1000 Oe magnetic field is presented in Fig. \ref{fig2}(a). The magnetization of the substrate was subtracted assuming temperature-independent diamagnetism. As the temperature decreases below 125 K, $M$ increases and a hysteresis behavior becomes visible around 110K. Below 100K, the measured $M$ is almost equal to the saturated magnetization. The inset of Fig.\ref{fig2}(a) shows the temporal behavior of $\varDelta T/T$ at temperatures of 54, 70 and 120 K and provides a direct experimental evidence of a relation between $\varDelta T_{\rm mag}/T$ and the demagnetization: just above the Curie temperature at 120 K, the slow response of the differential transmittance vanishes.

    The increase of the temperature ($\varDelta {\mathcal T}$) in the localized spin system can be quantitatively estimated if we assume that the lattice system is a constant-temperature heat bath with the total temperature rise of the spin system determined by the optical excitation power density and the lattice specific heat.\cite{Kojima} Given a temperature rise at a certain base temperature ${\mathcal T}_0$, the expected total variation of the magnetization
($\varDelta M_{\rm tot}$) due to photo-excitation can be calculated using the GaAs lattice-specific heat\cite{Blakemore} assuming that the small Mn fraction has a negligible effect. With the above interpretation, $\varDelta T_{\rm mag}/T$ should be proportional to $\varDelta M_{\rm tot}$ so that a fit of $\varDelta T_{\rm mag}/T$ and $\varDelta M_{\rm tot}$ can be performed using a scaling parameter. For an initial temperature of 54K, the temperature increase due to photo-induced heating is calculated to be around 40K.

    The circle and square data points in Fig. 3(b) show, on different linear scales, $\varDelta T_{\rm mag}/T$ as a function of ${\mathcal T}$ measured at 190 and 250 meV probe energies respectively. The solid curve is the differential of the magnetization calculated assuming a 660 ${\rm \mu J/cm^2}$ heat injection (i.e., for each data point, $\varDelta M_{\rm tot}$ is computed from the M vs ${\mathcal T}$ curve) and has a shape that closely matches both sets of data points. This value of the photo-induced temperature increase to a quasi-equilibrium state gives the best fit of the data points and is reasonable considering the energy of the pump pulse\cite{Kojima}.

    We describe next how results of the mid-infrared linear absorption and variation of $\varDelta T_{\rm mag}/T$ as a function of the probe energy support the ABMP model. Following the treatment of Lucovski\cite{Lucovski}, the energy dependent part of the linear infrared absorption due to a band to acceptor level transition can expressed as
\begin{equation}
\alpha(\hbar\omega, E_a)\propto\frac{(E_a)^{1/2}(\hbar\omega-E_a)^{3/2}}{(\hbar\omega)3}.
\label{absorption}
\end{equation}
The magnetic polaron formation effect can be considered by adding the exchange interaction energy between a hole and local spins $E_{\rm MP}$ to the acceptor binding energy $E_a$as schematically shown in Fig.\ref{fig3}(b).

\begin{figure}
\includegraphics[width=\linewidth]{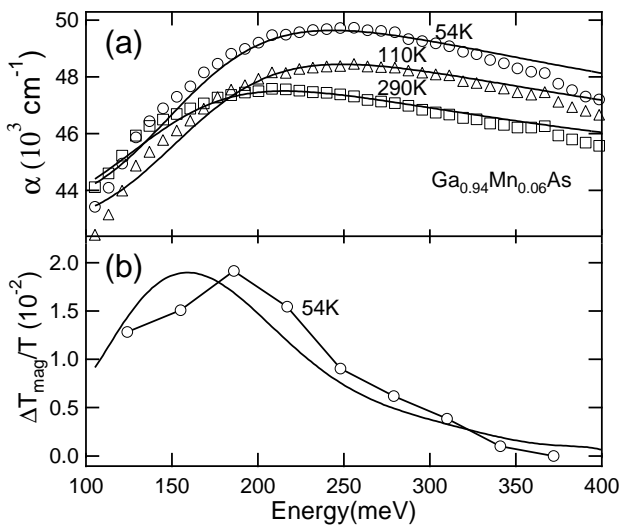}
\caption{
(a): Fitted (solid lines) and experimental (data points) curves of the linear absorption at  54, 110 and 290 K. (b)
Simulated (solid line) and experimental (data points) variation of $\varDelta T_{\rm mag}/T$ as a function of the
probe energy at 54 K.
}
\label{fig4}
\end{figure}

    Fig.\ref{fig4}(a) shows experimental values and simulation of the linear absorption at 54, 110 and 294 K. In the fitting, the fluctuation of $E_a$ due to the formation of random clusters commonly observed in heavily doped materials is taken into account\cite{Thomas} and a Thomas-Fermi density of states is also used to include band tailing
effects\cite{Kane}. For each experimental curve a proportionality constant is used as a second adjustable parameter to estimate the energy-independent part of the absorption. From the analysis, the total binding energy is found to reach
a maximum value around 129 meV at the Curie temperature and decrease to a value around 126 meV at 54K. At room temperature, the binding energy is 111 meV so that $E_{\rm MP}$(room temperature) $\approx$ 0 as expected. $E_{\rm
MP}$ is 15-18 meV around the Curie temperature. From this value, knowing that the doping ratio is 6 percent and assuming that the number of Mn ions around each hole is 5--6, the polaron radius can be estimated to be around 1
nm\cite{Lucovski}. The obtained value of $E_{\rm MP}$ is also comparable to the one obtained for phosphorous-doped Cd$_{1-x}$Mn$_x$Te\cite{Bugajski}, which is a material with comparable carrier-local spin exchange energy\cite{Okabayashi,Mizokawa} and acceptor binding energy\cite{Tarhan,Bugajski}.

    The amplitude of the slow response of the transmittance $\varDelta T_{\rm mag}/T$ shown by the data points in Fig. \ref{fig4}(b) provides strong evidence that a photo-excitation of the sample at 54 K results in an increase of the
polaron binding energy $E_{\rm MP}$. Using equation \eqref{absorption}, we easily obtain \begin{align}
\varDelta T_{\rm mag}/T&\approx-\varDelta\alpha l \notag\\
= -&[\alpha(\hbar\omega, E_a+E_{\rm MP}+\varDelta E_{\rm MP})-\alpha(\hbar\omega, E_a+E_{\rm MP})]l
\end{align} where $l$ is the thickness of the absorbing layer and $\varDelta E_{\rm MP}$ is the only fitting parameter. The solid curve in Fig. \ref{fig4}(b) shows the best fit that could be obtained with this simple expression, corresponding to a positive value $\varDelta E_{\rm MP}\approx$ 2meV. A slight discrepancy between the calculated curve and the experimental data is observed but the important point to emphasize is that $\varDelta E_{\rm MP}$ is positive (blueshift) and in the range of a few meV. Assuming a photo-induced change of 40K as in Fig. \ref{fig2}(b), this variation of the polaron binding energy is in overall agreement with the $E_{\rm MP}$ values obtained from the linear measurements shown in Fig. \ref{fig4}(a). Demagnetization leads to an increased polaron binding energy, strongly suggesting that ferromagnetism is due to a coupling between the local spins and localized holes.

    With the spectrum shown in Fig.\ref{fig4}(b), we succeeded in isolating the ferromagnetism-related component of the linear mid-infrared absorption peak. Compared to the raw absorption shown in Fig.\ref{fig4}(a) the line shape is
significantly shifted to lower energies, in qualitative agreement with the analysis of Hirakawa et al.\cite{Hirakawa}. However, the spectral features observed are different and a rising edge is seen in Fig.\ref{fig4}(b).
Results strongly suggest that the magnetic polarons play an important role in the ferromagnetism of Ga$_{1-x}$Mn$_x$As.

    The overall mechanism of photo-induced demagnetization can be described as follows. The photo-excited electrons undergo a rapid non-radiative recombination within a few picoseconds leading to a temperature increase of the holes
and the lattice. Due to the spin-dependent band structure of the holes, an increase of the kinetic energy does not directly affect the spin configuration. Thus in the first few picoseconds following photo-excitation the binding
energy of the magnetic polarons does not change. After a few tens of picoseconds, the spin temperature gradually increases to reach quasi-thermal equilibrium with the hole and lattice systems. The Mn spins not located within a
polaron radius loose their alignment more easily than the ones located within a polaron radius. As a result, the spatial variation of the magnetic potential in the material increases and the Mn-acceptor binding energy increases. In other words, the Mn acceptor level is shifted due to photo-induced demagnetization.

    In conclusion, we have shown that mid-infrared pump-probe transmittance characterization, a purely optical experimental technique, allows to observe ferromagnetic behavior in Ga$_{1-x}$Mn$_x$As due to the spin-related,
slow-varying part of the differential transmittance. The component of the mid-infrared absorption related to ferromagnetism could be extracted. The variation of the photo-induced demagnetization as a function of time, probe
energy and temperature can be explained with the ABMP model.

This work is partly supported by JSPS, KAKENHI (S) and MEXT, KAKENHI in the Priority Area ``Semiconductor
Nanospintronics".

\end{document}